\documentclass[3p,twocolumn]{elsarticle}
\pdfoutput=1
\DeclareGraphicsExtensions{.pdf}

\biboptions{sort&compress}
\usepackage[colorlinks=true,linkcolor=blue, citecolor=blue, urlcolor=blue]{hyperref}
\usepackage{natbib}
\usepackage{color}
\usepackage{multirow}
\usepackage{amssymb}
\usepackage{amsmath}
\usepackage{slashed}
\usepackage{graphicx}
\usepackage{placeins}
\usepackage{array}
\usepackage{dblfloatfix}

\newcommand{\PbPb}{\mbox{Pb+Pb}}

\newcommand{\jpsi}{\mbox{$J/\psi$}}
\newcommand{\psitwo}{\mbox{$\psi(2S)$}}

\newcommand{\ptini}{\mbox{$p_{T,\mathrm{ini}}$}}

\newcommand{\Npart}{\mbox{$N_{\mathrm{part}}$}}

\newcommand{\Raa}{\mbox{$R_{\rm AA}$}}

\newcommand{\Raaj}{\mbox{$R_{\rm AA}^{J/\psi}$}}
\newcommand{\Raap}{\mbox{$R_{\rm AA}^{\psi(2S)}$}}

\newcommand{\pt}{\mbox{$p_{T}$}}

\newcommand{\ptjet}{\mbox{$p_{T}^{\mathrm{jet}}$}}

\newcommand{\Rdz}{\mbox{$R_{D(z)}$}}

\newcommand{\cf}            {c_\mathrm{F}}

\begin{document}

\begin{frontmatter}
\title{
\center{Flavor aspects of parton energy loss}
}

\author[Charles]{Martin Spousta}
\address[Charles]{Institute of Particle and Nuclear Physics, Charles University, Prague, Czech Republic}

  \begin{abstract}
  Understanding flavor dependence of the parton energy loss is one of
key tasks of the jet quenching physics. In these proceedings we provide
a summary of recent works on a quantification of the flavor dependence
of parton energy loss along with a summary and discussion of a subset of
contributions presented at the Hard Probes 2016 conference which are
related to the flavor aspects of parton energy loss.
  \end{abstract}


\end{frontmatter}

\newlength{\fighalfwidth}
\setlength{\fighalfwidth}{0.49\textwidth}

\section{Introduction}

Understanding flavor dependence of the parton energy loss is one of
key tasks of the jet quenching physics. 
  The original results presented in these proceedings are largely based 
on Refs.~\cite{Spousta:2015fca,Spousta:2016agr} that put forward a model 
or method which allows to quantify some of the basic properties of the 
parton energy loss. Besides that, we included here also a summary and 
discussion of a subset of contributions presented at the conference 
which are related to the flavor aspects of the parton energy loss and we 
include also an outcome of some of discussions which took place at the 
conference.

This paper is organized as follows: in the first section we discuss 
basic features seen in the data on inclusive charged particles and jets 
in \PbPb\ collisions at the LHC and relate them with the flavor 
dependence of the parton energy loss.
  In the second section we put forward a quantification of the flavor 
dependence of the parton energy loss and provide further discussion on 
the impact of the flavor dependence on dijet measurements, jet 
substructure measurements, and measurements employing charged particles.
  In the last section we discuss the latest results on charmonia in the 
kinematic domain of large-\pt\ and we point to a similarity between the 
jet quenching and charmonia suppression which indicates that radiative 
energy loss may be a dominant source of the energy loss of charmonia at 
high-\pt.

\section{Suppression of inclusive charged particles and jets at the LHC}
\label{sec:one}

Many precise measurements of the suppression of inclusive charged 
particles quantified using the nuclear modification factor, \Raa, have 
been published by LHC experiments 
\cite{Abelev:2012hxa,CMS:2012aa,Aad:2015wga,Khachatryan:2016odn}. 
Direct quantification of the magnitude of parton energy loss using the 
charged particle \Raa\ is not straightforward since the correspondence 
between the kinematics of the initial parton and observed final state 
hadron is smeared by the fragmentation process. Consequently, more 
direct quantification of the parton energy loss may be done from the 
measurements of jet \Raa. At this conference a new result on the \Raa\ 
of fully reconstructed jets with different jet sizes has been presented 
\cite{Khachatryan:2016jfl}. This new measurement follows the previously 
measured jet \Raa\ \cite{Aad:2014bxa}. Complementary to these 
measurements is the measurement of fragmentation functions which was 
presented at this conference by several speakers and which was published 
in Refs.~\cite{Aad:2014wha,Chatrchyan:2014ava,ATLASFragPreliminary}.

  The above cited results bring questions about some interesting features seen in the data: 
  {\it 
  \begin{itemize}
  \item Why do have the jet \Raa\ and charge particle \Raa\ almost no rapidity 
dependence given different input parton spectra and flavor 
composition at different rapidities? 
  \item What is responsible for the 
enhancement at high momentum fractions ($z$) seen in the fragmentation? Can we 
find a connection among charged particle \Raa, jet \Raa\ and jet 
fragmentation measurement? 
  \item Having the jet \Raa\ at hand can we directly quantify the 
size of the parton energy loss?
  \end{itemize}
  }
  To answer these questions we introduced a model \cite{Spousta:2015fca} 
which is based on parameterizations of initial parton spectra and the 
parton energy loss. The only assumption on the physics of the jet 
quenching in this model is the functional form for the parton 
energy loss which is assumed to be of the power-law form -- the total 
transverse momentum lost by the parton is

\begin{equation}
 \Delta \pt = c_F \cdot s \cdot \bigg( \frac{\ptini }{ p_{T,0}} \bigg)^\alpha
\end{equation}
  Here $s$, $\alpha$, and $\cf$ are free parameters of 
the model, \ptini\ is the transverse momentum of a parton initiating a 
jet and $p_{T,0}$ is an arbitrary scale (set to 40~GeV).
  Parameter $\cf$ represents a color factor which quantifies the 
difference between the in-medium radiation of quark-initiated jets and 
gluon-initiated jets. 
  For the first studies, the $\cf$ was fixed to be 1 and 
$C_A/C_F = 9/4$ for light-quark-initiated jets and gluon-initiated jets, 
respectively.

  For a better orientation we label this our approach Effective Quenching (EQ) 
model but it could very well be called {\it the model independent method} which 
allows to extract basic properties of the average parton energy loss.
  
  The EQ model is capable of describing the full $\ptjet$, rapidity, and 
centrality dependence of the measured jet \Raa\ using three effective 
parameters which are obtained by minimizing with respect to the \Raa\ 
data published in Ref.~\cite{Aad:2014bxa}. The successful description of 
the jet \Raa\ in the full kinematic space implies that the absence of a clear 
rapidity dependence seen in the data comes from a cancellation between 
two competing effects which evolve with increasing rapidity: steepening 
of initial parton spectra and enhancing the fraction of quark initiated 
jets. While the former alone generally leads to a smaller \Raa\ the later 
alone generally leads to a larger $\Raa$.

  The quantification of the average parton energy loss provided by 
minimizing the difference between the model and the data 
revealed three interesting properties of the energy loss:
  1) the magnitude of the energy loss, $s$, depends linearly on the 
$\Npart$; 2) the power $\alpha$ is approximately 0.5 and it is constant 
as a function of $\Npart$; 3) the linear dependence of $s$ on $\Npart$ 
does not extrapolate to zero for $\Npart$ approaching zero. For more 
details on this quantification see 
Refs.~\cite{Spousta:2015fca,Spousta:2016agr}.

  The model can be further used to evaluate the impact of the change in 
the jet spectra on the measured fragmentation functions in a simple way. 
The main principle of the procedure is following: subtract the energy from the initial parton 
and then let it fragment as in the vacuum. The 
modifications of fragmenation functions were quantified in the data e.g. 
by a ratio, \Rdz, of fragmentation functions measured in central 
collisions to those measured in peripheral or proton-proton collisions. 
The modifications seen from \Rdz, excluding the enhancement at low-$z$, 
are described by the model. Thus, it may be concluded that these 
modifications result primarily from the different quenching of the 
quark and gluon jets. The assumption on the fragmentation of the 
quenched parton used here reflects a physics scenario in which the 
parton shower looses the energy coherently. Indeed, it was recognized in 
several theoretical papers that such color coherence effects play an 
important role in the jet quenching process 
\cite{MehtarTani:2010ma,MehtarTani:2011tz,CasalderreySolana:2012ef,Blaizot:2013hx}. 
The successful description of the jet fragmenation when employing this 
assumption within EQ model may be considered an independent argument
speaking in favor of the physics scenario based on the color coherence.

  The charged particle \Raa\ and jet \Raa\ can in principle be connected 
using fragmentation functions since each charged particle with 
sufficiently high-\pt\ which does not come from the underlying event has 
to be found in a jet. The fact that the model can reasonably well 
reproduce the \Raa\ of inclusive charged particles at $\pt \gtrsim 
20$~GeV is a cross-check. Besides that, it answers a question which was 
posed several times at this conference: {\it How to reconcile the fact 
that the charged particle \Raa\ reaches values greater than the values 
of \Raa\ of inclusive jets?} The answer to that question is that such a 
direct interpretation of the charged particle \Raa\ is not possible 
since charged particle \Raa\ is a non-trivial convolution of flavor 
dependent jet suppression and fragmentation functions. The fact that the 
EQ model can reproduce all three kinds of jet related measurements 
implies that the data do not contradict each other.

  It should be mentioned that the advantage of the above described modelling 
is that it allows to obtain exact analytic formulae for the \Raa\ and 
\Rdz\ which can then be used in a straightforward way to extract 
information about the modification of jet yields and jet structure.
  Another advantage is that the model employs {\it minimal} assumptions 
on the physics of the jet quenching. While this may be judged as an 
advantage it may also be judged as a disadvantage since we explore here 
just the {\it average} jet quenching ignoring e.g. the path-length 
dependence of the quenching or the role of fluctuations in the jet 
quenching. On the other hand, if the model fails in describing some 
observable, it allows to quantify the impact of these effects on that 
observable as discussed in the next section.

  There are several predictions that can be made based on this 
modelling: 
  \begin{enumerate}
  \item Jet \Raa\ should start to decrease at forward rapidities. 
  \item If the magnitude of the energy loss is the same in 2.76 TeV and 5.02 TeV \PbPb\ collisions, then the jet \Raa\ measured at 5.02 TeV should be very similar to 
the \Raa\ measured at 2.76~TeV.
  \item The enhancement observed in \Rdz\ 
distributions for fragments with high-$z$ should be smaller for more 
forward jets.
  \item \Rdz\ distributions for fragments with $z$ approaching 
one should start to decrease. 
  \end{enumerate}
  In particular prediction 3) and 4) appears to be seen in the new precise
data shown at this conference \cite{ATLASFragPreliminary}.

\section{Quantifying the role of the flavor}

  Given the quantification of the 
energy loss of inclusive jets provided in the previous section, more 
questions can be asked: {\it 
  \begin{enumerate}
  \item Can we quantify a difference between the 
jet quenching of light-quark jets and b-jets? 
  \item Can we quantify the 
color-charge dependence of the jet quenching? 
  \item What observables 
are sensitive to the difference between the quenching of quark and 
gluon initiated jets?
  \end{enumerate}
  }

\begin{figure*}[h]
\begin{center}
\includegraphics[width=0.9\textwidth]{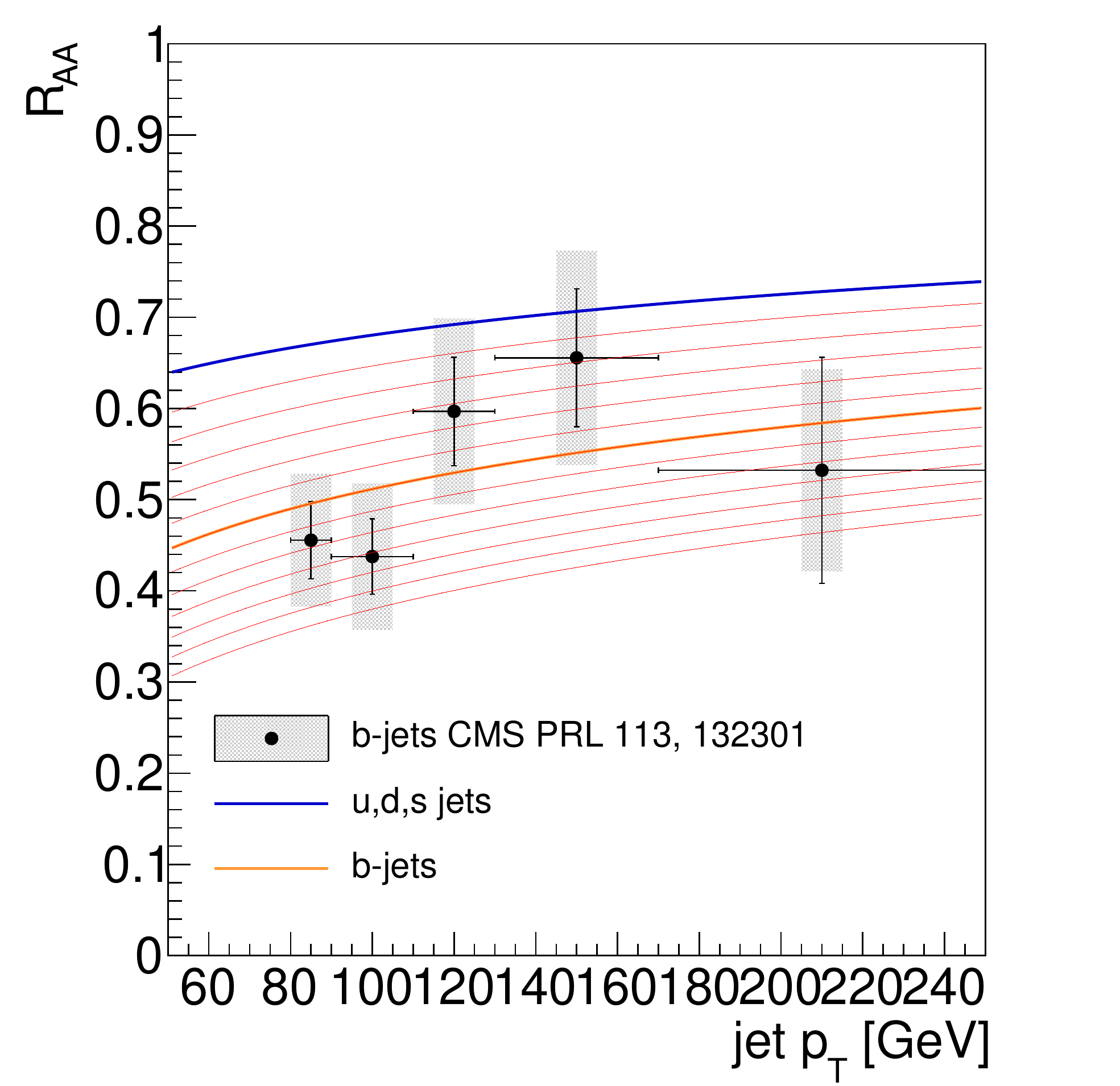}
\end{center}
\caption{
  Evaluation of the difference between the quenching of light-quark 
initiated jets and b-quark initiated jets within the EQ model. Blue line 
represents the energy loss of b-jets for the case that they are 
suppressed by the same magnitude as the light-quark initiated jets. 
Orange lines represent multiples of light quark suppression which are 
inputs to the determination of the minimal difference between the 
simulation and the data. Thick orange line represents a result of the 
minimization (for details see the text).
  }
\label{fig:fig1}
\end{figure*}

  To answer the first question we can compare the b-jet quenching 
simulation with the data from Ref.~\cite{Chatrchyan:2013exa}. The b-jet quenching 
simulation used the EQ model which was first run with the assumption 
that the b-jet energy loss is the same as the energy loss of light quark 
jets. Then energy loss was increased in steps and the result was 
compared with the data as shown in Fig.~\ref{fig:fig1}. By minimizing the difference between the data 
and the simulation, it was found that the b-jets are quenched $1.5 \pm 
0.4$ times more than the light-quark jets. The evaluation of this 
relative suppression assumes that $30\% \pm 10\%$ of b-jets are produced 
in the gluon splitting \cite{Chatrchyan:2013exa} for which the gluon jet suppression is used. 

  The second question seeking for the quantification of the difference 
between the in-medium radiation of quark and gluon jets is one of the 
basic questions of the jet quenching physics. The knowledge of the 
color-charge dependence of the parton energy loss should provide a direct 
link between the in-medium parton radiation and basics of the pQCD 
represented at the leading order by SU(3) Casimir invariants $C_A$ and 
$C_F$.
  Attempts and proposals were put forward to extract that difference using 
identified charged particles in past, see e.g. 
Refs.~\cite{Armesto:2005iq,Mohanty:2007dv}. However, no direct extraction 
of the color factor in heavy ion environment existed as of now. To 
extract the color factor, more precise quantification of the quenching 
of inclusive jets was done using a global fit and the modeling of the 
initial parton spectra using precise NLO description as described in 
~\cite{Spousta:2016agr}. This allowed to extract the color factor which quantifies the 
difference between the in-medium radiation of quark-initiated jets and 
gluon-initiated jets. The value $\cf = 1.78 \pm 0.12$ was obtained. This 
value is consistent with the value calculated and measured in the vacuum 
which is $\approx 1.8$ at the jet hardness $Q = 100$~GeV 
\cite{Capella:1999ms,Acosta:2004js}. 

  While the information about the color-charge dependence of the jet 
quenching is interesting by itself, the fact that quark and gluon 
initiated jets are suppressed differently implies that certain 
observables aiming to quantify modifications of the jet substructure or a 
path-length dependence of the jet quenching can be biased by this basic 
feature of the parton energy loss. Specifically at this conference two 
new measurements were discussed: the measurement of the splitting 
distribution $z_g$ \cite{CMS:2016jys} and the measurement of the dijet 
asymmetry distribution \cite{ATLASdijetPreliminary}. The former 
measurement can be used to access the modifications of the jet internal 
structure at the level of subjets, the later measurement can be used to 
quantify e.g. the path-length dependence of the jet quenching or to access the 
role of the fluctuations in the jet quenching. By employing the EQ model 
and checking the behaviour of observables in POWHEG+PYTHIA8 \cite{Alioli:2010xd,Alioli:2010xa,Sjostrand:2014zea} 
simulation, it was concluded that the $z_g$ distributions are not 
sensitive to flavor of initial partons and that asymmetry distributions 
exhibit only a modest sensitivity to the flavor.
  To build a basic knowledge about sensitivity of the dijet asymmetry to 
processes that go beyond the average energy loss we use the EQ model to 
calculate a difference between the leading and subleading jet quenching 
which is needed to reproduce the data.
  It was found that in order to reproduce the maximum seen in the 
measured asymmetry distribution in 0-10\% central collisions for leading 
jets with $\pt = 100-126$~GeV, the subleading jet needs to be quenched 
three times more on average than the leading jet.

  The conclusion about no or weak sensitivity of the splitting and dijet 
asymmetry distributions on the initial parton flavor does not obviously 
hold e.g. for the \pt\ spectra of inclusive charged particles which are 
emitted from different partons at different transverse momenta or center 
of mass energies. This is one aspect which is more or less obvious but 
important and which needs to be taken into account e.g. when comparing 
the magnitude of inclusive charged particle \Raa\ at RHIC and at the 
LHC. Another aspect which is also more or less obvious and which needs 
to be considered for such comparisons is the difference in the slopes 
(and shapes) of initial parton spectra. This may be put into the context 
of the discussion about the puzzling difference of the size of K-factors 
quantifying the magnitude of the jet quenching which appear to be larger 
at RHIC than at the LHC, see Ref.~\cite{Andres:2016iys}, which was also 
briefly discussed at this conference. Indeed, when using precise NLO 
initial parton spectra for RHIC and LHC energies along with the same 
magnitude of the parton energy loss as an input to the EQ model, one 
obtains larger values of \Raa\ for LHC energies than for RHIC energies 
which might be interpreted at a first glance as a larger quenching at 
RHIC than at the LHC. The judgment about these biases which might appear 
in some studies of the jet quenching are left for respective authors. 
The discussion presented here only aims to draw the attention to these 
flavor related aspects of the jet quenching which calls for an explicit 
treatment (see e.g. Ref.~\cite{Wang:2016fds}).

\section{Radiative energy loss of charmonia}

   The ability of EQ model to successfully describe jet \Raa\ in all 
measured rapidity bins, inclusive charged particle \Raa\ at high-\pt, and 
details seen in the inclusive jet fragmentation functions, which was 
discussed in the Sec.~\ref{sec:one}, speaks strongly in favor of a 
physics picture in which parton shower or its large part looses the 
energy coherently.
    If the parton shower, or its large part, radiates as one object, one 
can ask if it is possible to find some similarities between the 
suppression of jets and a suppression of other objects with an internal 
structure.
  One such candidate are the charmonia. 
  The similarity between the jet suppression and charmonia suppression 
was explored in Ref.~\cite{Spousta:2016agr} and it is briefly summarized here in the 
context of other results presented at this conference.

  There is no unique interpretation of the charmonia suppression 
measurements as of now \cite{Andronic:2015wma}. New measurements at the 
LHC 
\cite{Aad:2010aa,Chatrchyan:2012np,Abelev:2013ila,Khachatryan:2014bva,Adam:2015rba,CMS:2012vxa}
  should provide more insight to the mechanism of the charmonia 
suppression. The precise measurements of the prompt \jpsi\ in the muon 
channel \cite{Chatrchyan:2012np,CMS:2012vxa} showed that the nuclear 
modification factor, \Raaj , reaches a value of $\sim 0.2$ in the most 
central collisions ($\Npart \gtrsim 350$), continuously grows up to a 
value of $\sim 0.6-0.7$ reached in the most peripheral collisions 
($\Npart \lesssim 50$). The \Raaj\ exhibits only a weak (if any) 
dependence on the \jpsi\ momentum in the region of pi $\pt = 6.5 - 
30$~GeV and $|y|<2.4$. The dependence of \Raaj\ on the rapidity is also 
weak. More recently, a prompt production of \psitwo\ was also measured 
in terms of a double ratio of measured yields, $(N_{\psi(2S)} / 
N_{J/\psi})|_\mathrm{Pb+Pb} / (N_{\psi(2S)} / N_{J/\psi})|_{pp} = \Raap 
/ \Raaj$ \cite{Khachatryan:2014bva}. It was shown that \psitwo\ yields 
are suppressed by a factor of $\sim 2$ with respect to \jpsi\ in the 
range $|y|<1.6$ and $6.5 < \pt < 30$~GeV. At this conference, also new 
results were shown on the charmonia suppression in 5.02~TeV Pb+Pb data 
measured by ATLAS \cite{ATLAS:2016qpn} and CMS \cite{Sirunyan:2016znt} 
where quantitatively similar trends 
in the charmonia suppression 
to those in 2.76~TeV data
were observed.

To test the idea of similarity in the physics of jet quenching and 
prompt charmonia suppression the EQ model 
has been employed. The input to the model were \pt\ spectra of 
\jpsi\ and \psitwo\ and effective parameters obtained from the analysis 
of jet \Raa . The \pt\ spectra were obtained from PYTHIA8 
which was 
reweighted to reproduce the data measured in $pp$ collisions at the 2.76 
TeV \cite{ATLAS:2015pua}.
  The realistic $\pt$ spectra were then used as an input to the EQ model
which was run with two different settings of the color factor: first,
corresponding to the color factor for the energy loss of light-quark
initiated jets (defined to be one), and second, corresponding to the
color factor extracted for the energy loss of gluon-initiated jets.
  A very good agreement of the model with the data for the case of the
light-quark energy loss was seen. The \Npart\ dependence of \Raaj , its
\pt\ and rapidity dependence were reproduced.
  Remarkably, the model was also able to very well reproduce the 
suppression of \psitwo\ from Ref.~\cite{Khachatryan:2014bva} which was 
quantified in terms of the ratio of nuclear modification factors, $\Raap 
/ \Raaj$.
  The striking similarity between the measured \jpsi\ and \psitwo\
suppression and the energy loss of jets suggests that the radiative energy
loss may be a dominant contribution to the energy loss of charmonia in the
studied kinematic region.

  Quantification of the magnitude of the parton energy loss and the role 
of the flavor in the parton energy loss along with the observations of 
the similarity between the jet quenching and charmonia suppression 
should improve the understanding of physics mechanism behind both, the 
jet quenching and charmonia suppression.

\section*{Acknowledgment}
   This work was supported by Charles University Research Development
Scheme (PRVOUK) P45, UNCE 204020/2012 and MSMT grant INGO II LG15052.

\bibliographystyle{elsarticle-num}
\bibliography{forArchiv}

\end{document}